\documentclass[dvips]{article}
\usepackage{supertabular,lscape,epsfig}
\usepackage{amssymb}
\usepackage{amsmath}

\DeclareSymbolFont{ppa}{OT1}{ppl}{m}{it}
\DeclareMathSymbol{\vv}{\mathalpha}{ppa}{'166}

\begin{document}

\newcommand{\dd}{\,{\rm d}}
\newcommand{\ie}{{\it i.e.},\,}
\newcommand{\etal}{{\it et al.\ }}
\newcommand{\eg}{{\it e.g.},\,}
\newcommand{\cf}{{\it cf.\ }}
\newcommand{\vs}{{\it vs.\ }}
\newcommand{\zdot}{\makebox[0pt][l]{.}}
\newcommand{\up}[1]{\ifmmode^{\rm #1}\else$^{\rm #1}$\fi}
\newcommand{\dn}[1]{\ifmmode_{\rm #1}\else$_{\rm #1}$\fi}
\newcommand{\upd}{\up{d}}
\newcommand{\uph}{\up{h}}
\newcommand{\upm}{\up{m}}
\newcommand{\ups}{\up{s}}
\newcommand{\arcd}{\ifmmode^{\circ}\else$^{\circ}$\fi}
\newcommand{\arcm}{\ifmmode{'}\else$'$\fi}
\newcommand{\arcs}{\ifmmode{''}\else$''$\fi}
\newcommand{\MS}{{\rm M}\ifmmode_{\odot}\else$_{\odot}$\fi}
\newcommand{\RS}{{\rm R}\ifmmode_{\odot}\else$_{\odot}$\fi}
\newcommand{\LS}{{\rm L}\ifmmode_{\odot}\else$_{\odot}$\fi}

\newcommand{\Abstract}[2]{{\footnotesize\begin{center}ABSTRACT\end{center}
\vspace{1mm}\par#1\par
\noindent
{~}{\it #2}}}

\newcommand{\TabCap}[2]{\begin{center}\parbox[t]{#1}{\begin{center}
  \small {\spaceskip 2pt plus 1pt minus 1pt T a b l e}
  \refstepcounter{table}\thetable \\[2mm]
  \footnotesize #2 \end{center}}\end{center}}

\newcommand{\TableSep}[2]{\begin{table}[p]\vspace{#1}
\TabCap{#2}\end{table}}

\newcommand{\FigCap}[1]{\footnotesize\par\noindent Fig.\  %
  \refstepcounter{figure}\thefigure. #1\par}

\newcommand{\TableFont}{\footnotesize}
\newcommand{\TableFontIt}{\ttit}
\newcommand{\SetTableFont}[1]{\renewcommand{\TableFont}{#1}}

\newcommand{\MakeTable}[4]{\begin{table}[htb]\TabCap{#2}{#3}
  \begin{center} \TableFont \begin{tabular}{#1} #4 
  \end{tabular}\end{center}\end{table}}

\newcommand{\MakeTableSep}[4]{\begin{table}[p]\TabCap{#2}{#3}
  \begin{center} \TableFont \begin{tabular}{#1} #4 
  \end{tabular}\end{center}\end{table}}

\newenvironment{references}%
{
\footnotesize \frenchspacing
\renewcommand{\thesection}{}
\renewcommand{\in}{{\rm in }}
\renewcommand{\AA}{Astron.\ Astrophys.}
\newcommand{\AAS}{Astron.~Astrophys.~Suppl.~Ser.}
\newcommand{\ApJ}{Astrophys.\ J.}
\newcommand{\ApJS}{Astrophys.\ J.~Suppl.~Ser.}
\newcommand{\ApJL}{Astrophys.\ J.~Letters}
\newcommand{\AJ}{Astron.\ J.}
\newcommand{\IBVS}{IBVS}
\newcommand{\PASP}{P.A.S.P.}
\newcommand{\Acta}{Acta Astron.}
\newcommand{\MNRAS}{MNRAS}
\renewcommand{\and}{{\rm and }}
\section{{\rm REFERENCES}}
\sloppy \hyphenpenalty10000
\begin{list}{}{\leftmargin1cm\listparindent-1cm
\itemindent\listparindent\parsep0pt\itemsep0pt}}%
{\end{list}\vspace{2mm}}

\def\TYLDA{~}
\newlength{\DW}
\settowidth{\DW}{0}
\newcommand{\dw}{\hspace{\DW}}

\newcommand{\refitem}[5]{\item[]{#1} #2%
\def\REFARG{#3}\ifx\REFARG\TYLDA\else, {\it#3}\fi
\def\REFARG{#4}\ifx\REFARG\TYLDA\else, {\bf#4}\fi
\def\REFARG{#5}\ifx\REFARG\TYLDA\else, {#5}\fi.}

\newcommand{\Section}[1]{\section{\hskip-6mm.\hskip3mm#1}}
\newcommand{\Subsection}[1]{\subsection{#1}}
\newcommand{\Acknow}[1]{\par\vspace{5mm}{\bf Acknowledgements.} #1}
\pagestyle{myheadings}

\newfont{\bb}{ptmbi8t at 12pt}
\newcommand{\xrule}{\rule{0pt}{2.5ex}}
\newcommand{\xxrule}{\rule[-1.8ex]{0pt}{4.5ex}}
\def\thefootnote{\fnsymbol{footnote}}
\begin{center}
{\Large\bf The Optical Gravitational Lensing Experiment.
\vskip1pt
Search for Planetary and Low-Luminosity 
\vskip1pt
Object Transits in the Galactic Disk.
\vskip3pt
Results of 2001 Campaign -- Supplement\footnote{Based on observations
obtained with the 1.3~m Warsaw telescope at the Las Campanas Observatory
of the Carnegie Institution of Washington.}}
\vskip.6cm
{\bf A.~~U~d~a~l~s~k~i$^1$,~~K.~~\.Z~e~b~r~u~\'n$^1$, ~~M.~~S~z~y~m~a~{\'n}~s~k~i$^1$,
~~M.~~K~u~b~i~a~k$^1$, ~~I.~~S~o~s~z~y~\'n~s~k~i$^1$,
~~O.~~S~z~e~w~c~z~y~k$^1$, ~~\L.~~W~y~r~z~y~k~o~w~s~k~i$^1$,
~~and~~G.~~P~i~e~t~r~z~y~\'n~s~k~i$^{2,1}$}
\vskip2mm
$^1$Warsaw University Observatory, Al.~Ujazdowskie~4, 00-478~Warszawa, Poland\\
e-mail: (udalski,zebrun,msz,mk,soszynsk,szewczyk,wyrzykow,pietrzyn)@astrouw.edu.pl\\
$^2$ Universidad de Concepci{\'o}n, Departamento de Fisica,
Casilla 160--C, Concepci{\'o}n, Chile
\end{center}

\vspace*{7pt}  
\Abstract{The photometric data collected during 2001 season OGLE-III 
planetary/low luminosity object transit campaign were reanalyzed with the new 
transit search technique -- the BLS method by Kovacs, Zucker and Mazeh. In 
addition to all presented in our original paper transits, additional 13 
objects with transiting low-luminosity companions were discovered. We present 
here a supplement to our original catalog -- the photometric data, light 
curves and finding charts of all 13 new objects. 

The model fits to the transit light curves indicate that a few new objects may 
be Jupiter-sized (${R<1.6~R_{\rm Jup}}$). OGLE-TR-56 is a particularly 
interesting case. Its transit has only 13~mmag depth, short duration and a 
period of 1.21190 days. Model fit indicates that the companion may be 
Saturn-sized if the passage were central. 

Spectroscopic follow-up observations are encouraged for final classification 
of the transiting objects as planets, brown dwarfs or late M-type dwarf stars. 
 
We also provide the most recent ephemerides of other most promising planetary 
transits: OGLE-TR-10 and OGLE-TR-40 based on observations collected in June 
2002. 

All photometric data are available to the astronomical community from the OGLE 
Internet archive.}

\Section{Introduction}
In the paper by Udalski \etal (2002), the OGLE photometric survey presented 
results of the pilot campaign aiming at discovery of small depth transits 
caused by planets or low-luminosity small objects like brown dwarfs or late 
type M-dwarfs. The campaign turned out to be very successful -- 46 objects 
with transiting objects causing small (${<0.08}$~mag) box-shaped drop of 
brightness were detected. Simple transit model assuming completely dark 
transiting object and limb darkening of the primary star was fitted to the 
light curves of transits, indicating small sizes of transiting companions. In 
several cases the results indicated Jupiter-sized objects (${R<1.6~R_{\rm 
Jup}}$). 

Unfortunately, the  photometry alone cannot unambiguously distinguish between 
Jupiter size planets and other low-luminosity objects: brown dwarfs and late 
type M dwarfs. All of them have radii of the order of 0.1--0.2~\RS\ 
(${1{-}2~R_{\rm Jup}}$). Thus, a spectroscopic follow-up and a measurement of 
the radial velocity amplitude of the stars is needed to determine the masses 
of transiting companions and final classification. 

Large number of discovered objects with transiting companions makes the 
transit method of planetary search very attractive as the efficiency of search 
can be much higher than that of standard spectroscopic searches. Moreover, 
non-planetary companions like brown dwarfs or faint late M-type dwarf are also 
very interesting astronomically and their parameters are poorly known. It 
should be noted that photometry of transits combined with precise spectroscopy 
is the only method of unambiguous determination of all basic parameters like 
dimensions and masses. 
 
Recently, Kovacs, Zucker and Mazeh (2002) proposed a new method of analysis of 
large photometric datasets for transit detection: the Box-fitting Least 
Squares (BLS) algorithm. Encouraged by promising simulations on artificial 
datasets presented by Kovacs \etal (2002) we ran the method on photometric 
data of all objects presented in Udalski \etal (2002). All of them were easily 
found by the BLS method. Encouraged even more, we decided to run the algorithm 
on photometric data of all ${\approx 52~000}$ Galactic disk stars selected for 
transit search. Results were very impressive -- 13 additional objects with 
transiting low-luminosity companions were detected by the BLS method, 
increasing the total number of transit objects detected in the 2001 OGLE-III 
campaign data to 59. 

This paper is a supplement to the original paper presenting results of the 
2001 OGLE-III planetary/low-luminosity object transit campaign (Udalski \etal 
2002). We show here photometry, light curves and finding charts as well as 
results of preliminary analysis of transit light curves for all 13 new objects 
detected with the BLS method. Photometry of new OGLE-III stars  with 
transits is available to the astronomical community from the OGLE Internet 
archive. 

\Section{Observational Data}
Observations were collected with the 1.3-m Warsaw telescope at the Las 
Campanas Observatory, Chile, (operated by the Carnegie Institution of 
Washington) equipped with a large field CCD mosaic camera, consisting of 
eight ${2048\times4096}$ pixel SITe ST002A detectors. The pixel size of each 
of the detectors is 15~$\mu$m giving the 0.26 arcsec/pixel scale at the focus 
of the Warsaw telescope. Full field of view of the camera is about 
$35\times 35.5$~arcmins. The gain of each chip is adjusted to be about 
1.3~e$^-$/ADU with the readout noise from about 6 to 9~e$^-$, depending on the 
chip. 

The photometric data were obtained  during 32 nights spanning 45 days from 
June~12, 2001. Three fields located in the densest stellar regions in the 
direction of the Galactic center were monitored continuously up to 35 times 
per night.  After that period single observations of these fields were done 
once every few nights till the end of 2001 Galactic bulge season in 
October~2001. Almost all observations were made in the {\it I}-band filter. 
The exposure time of each image was set to 120 seconds. Altogether about 800 
epochs were collected for each field during the 2001 season. 

\Section{Search for Transits with the BLS Method}
About 52~000 Galactic disk stars with accuracy of photometry ({\it rms}) 
better than 0.015~mag were selected from about 5 million stars observed in 
three fields  during the 2001 OGLE-III campaign. Details of the selection 
procedure are described in Udalski \etal (2002). All these objects were 
subject to the BLS algorithm for transit search. 

We wrote the BLS code procedure based on the description and exemplary Fortran 
code provided by Kovacs \etal (2002). We used the  algorithm with 200 phase 
bins, and the BLS statistics was calculated for frequencies from 0.1 [1/day] 
(${P=10}$~days) to 0.95 [1/day] (${P=1.053}$~days) with the step of 0.0001 
[1/day]. Transits of duration from 0.015 to 0.15 in phase were searched for. 

Only objects for which the highest peak in the BLS power spectrum (BLS 
statistics \vs frequency) exceeded the effective signal-to-noise parameter 
${\alpha=9.0}$ and signal detection efficiency ${{\rm SDE}=3.0}$ (see 
definitions in Kovacs \etal 2002) were further analyzed. However, for the 
lowest signal-to-noise detection we required proportionally larger SDE value 
to avoid too many false detections. Linear relation $\alpha$ \vs SDE  serving 
as a detection threshold was experimentally established   based on results of 
the BLS run on the transit sample from Udalski \etal (2002). 

It should be noted that the BLS method is very efficient numerically. It took 
only several CPU hours to run all 52~000 stars through the BLS algorithm 
compared to weeks with the method used in the original search by Udalski \etal 
(2002). 

\Section{Results of the BLS Search}
Beside of all transit objects detected in Udalski \etal (2002), the BLS search 
detected 13 additional objects with low-luminosity transiting objects. 
Actually, the number of new detections was larger. However, on the final list 
we only left  the cases that light curves indicate transit with high 
probability. Objects with shallow but clearly triangle-shaped eclipses were 
removed as they are probably grazing eclipses of normal stars. Also objects 
with indications of secondary eclipses -- very likely, regular eclipsing stars 
blended with brighter disk stars were not included. 

We present new transit objects in identical form as for the original 
detections: in the form of a catalog and atlas of light curves. Table~1 
contains basic data for each star: abbreviation in the form OGLE-TR-NN, 
equatorial coordinates (J2000.0 epoch), orbital period, epoch of mid-eclipse, 
{\it I}-band magnitude and an approximate ${V-I}$ color outside eclipse, the 
depth of eclipse, number of transits observed ($N_{\rm tr}$), and remarks. 
OII acronym in the remarks column indicates that the object was observed in 
the OGLE-II phase of the OGLE survey. For these stars both -- the mean {\it 
I}-band magnitude and ${V-I}$ color from OGLE-II photometric maps (Udalski 
\etal 2002, in preparation) were used to set the zero point of photometry, and 
both values are accurate to better than 0.05~mag. In the remaining cases the 
accuracy of the magnitude scale and ${V-I}$ color is not better than 0.1~mag. 
It should be also noted that in the  case  of OGLE-TR-56 its ephemeris was 
updated based on observations collected in 2002 observing season up to June 
2002. 

\MakeTable{l@{\hspace{6pt}}
c@{\hspace{5pt}}c@{\hspace{6pt}}c@{\hspace{5pt}}c@{\hspace{4pt}}
c@{\hspace{4pt}}c@{\hspace{4pt}}c@{\hspace{4pt}}c@{\hspace{4pt}}l}{12.5cm} {OGLE-III
planetary and low luminosity object transits} {\hline
\noalign{\vskip4pt} Name       & RA (J2000)  & DEC (J2000) &   $P$    &
$T_0$         & $I$    &$V-I$ &$\Delta I$  &  $N_{\rm tr}$ &Rem.\\
           &             &             & [days]   & --2452000     & [mag]  &[mag] &[mag]  &\\ 
\noalign{\vskip4pt}
\hline
\noalign{\vskip4pt}
OGLE-TR-47 & 17\uph52\upm15\zdot\ups53 & $-30\arcd13\arcm54\zdot\arcs0$ &   2.33590 & 76.45988 & 15.602 & 1.26 & 0.019 & 7 & OII \\ 
OGLE-TR-48 & 17\uph51\upm17\zdot\ups11 & $-30\arcd03\arcm01\zdot\arcs3$ &   7.22550 & 74.19693 & 14.770 & 1.10 & 0.022 & 2 & \\ 
OGLE-TR-49 & 17\uph51\upm14\zdot\ups41 & $-29\arcd54\arcm23\zdot\arcs8$ &   2.69042 & 75.53363 & 16.180 & 1.72 & 0.034 & 2 & \\ 
OGLE-TR-50 & 17\uph53\upm48\zdot\ups08 & $-29\arcd56\arcm00\zdot\arcs9$ &   2.24861 & 76.26084 & 15.923 & 1.48 & 0.048 & 3 & OII \\ 
OGLE-TR-51 & 17\uph52\upm36\zdot\ups00 & $-29\arcd37\arcm28\zdot\arcs9$ &   1.74832 & 74.99241 & 16.714 & 1.42 & 0.030 & 4 & OII \\ 
OGLE-TR-52 & 17\uph52\upm54\zdot\ups01 & $-29\arcd46\arcm33\zdot\arcs8$ &   1.32565 & 73.84778 & 15.593 & 1.41 & 0.034 & 8 & OII \\ 
OGLE-TR-53 & 17\uph53\upm09\zdot\ups82 & $-30\arcd06\arcm29\zdot\arcs6$ &   2.90565 & 77.59653 & 15.998 & 1.13 & 0.034 & 4 & OII \\ 
OGLE-TR-54 & 17\uph52\upm57\zdot\ups52 & $-30\arcd05\arcm33\zdot\arcs1$ &   8.16256 & 74.73849 & 16.473 & 1.82 & 0.026 & 3 & OII \\ 
OGLE-TR-55 & 17\uph57\upm28\zdot\ups50 & $-29\arcd43\arcm50\zdot\arcs4$ &   3.18473 & 77.04692 & 15.800 & 1.48 & 0.019 & 4 & \\ 
OGLE-TR-56 & 17\uph56\upm35\zdot\ups51 & $-29\arcd32\arcm21\zdot\arcs2$ &   1.21190 & 72.68492 & 15.300 & 1.26 & 0.013 &11 & \\ 
OGLE-TR-57 & 17\uph55\upm53\zdot\ups20 & $-29\arcd22\arcm28\zdot\arcs7$ &   1.67483 & 75.46493 & 15.698 & 1.51 & 0.026 & 5 & OII \\ 
OGLE-TR-58 & 17\uph55\upm08\zdot\ups27 & $-29\arcd48\arcm51\zdot\arcs3$ &   4.34244 & 76.81982 & 14.751 & 1.20 & 0.030 & 2 & OII \\ 
OGLE-TR-59 & 17\uph55\upm03\zdot\ups30 & $-29\arcd48\arcm48\zdot\arcs4$ &   1.49709 & 73.15416 & 15.196 & 1.22 & 0.026 & 9 & OII \\ 
\hline}

Light curves in graphical form are presented in Appendix: full phased 
light curve and close up around the eclipse. Additionally we also provide a 
finding chart -- ${60\times60}$ arcsec subframe of the {\it I}-band reference
image centered on the star. North is up and East to the left in these images. 

\Section{Discussion}
\vspace*{12pt}
The BLS method of transit detection (Kovacs \etal 2002) proved to be a very 
effective and efficient tool for analysis of huge photometric datasets for 
periodic transits. Application of the method to the real photometric data 
collected during the 2001 OGLE-III transit campaign allowed to increase 
significantly the sample of low-luminosity transit objects, which is now totaling 
at 59. It is worth noting that the BLS method detects transit with smaller 
depth -- the smallest depth in the case of the OGLE-TR-56 transit is only 13 
mmag -- and also transits in more noisy data (fainter objects). 

For all new transit objects we performed an analysis of transit light curve 
similar as in our original paper (Udalski \etal 2002). We estimated sizes of 
transiting objects by modeling their light curves assuming a completely dark 
companion and using formulae provided by Sackett (1999). These were 
numerically integrated at the appropriate phase points to produce model light 
curves. Details can be found in Udalski \etal (2002). 

The smallest size of the star and its companion is obtained when the transit 
is central, \ie for the inclination ${i=90\arcd}$. Non-central passage 
(smaller $i$) requires a larger size of both, the star and its companion. 
Table~2 lists the minimum size of the transiting companion and the 
corresponding size of the star (assuming ${M_s=1~\MS}$) to provide information 
on the expected dimensions of the transiting objects. It should be remembered 
that semi-major axis of the orbit scales as $ M^{1/3}\!$, and therefore the same 
scaling is appropriate for the sizes of stars and companions listed in 
Table~2. In the close-up windows in the Appendix we show model light curves 
for central transit (${i=90\arcd}$) drawn with a continuous bold line. 

\MakeTable{lcc}{12.5cm}{Dimensions of stars and companions for central
passage ($M_s=1~\MS$)}
{
\hline
\noalign{\vskip3pt}
Name       & $R_s$   & $R_c$ \\
&[\RS]&[\RS]\\
\noalign{\vskip3pt}
\hline
\noalign{\vskip3pt}
OGLE-TR-47 &    1.95 &    0.23 \\
OGLE-TR-48 &    1.21 &    0.16 \\
OGLE-TR-49 &    0.97 &    0.15 \\
OGLE-TR-50 &    1.16 &    0.22 \\
OGLE-TR-51 &    0.85 &    0.13 \\
OGLE-TR-52 &    1.27 &    0.20 \\
OGLE-TR-53 &    0.76 &    0.12 \\
OGLE-TR-54 &    1.38 &    0.19 \\
OGLE-TR-55 &    0.93 &    0.11 \\
OGLE-TR-56 &    0.80 &    0.08 \\
OGLE-TR-57 &    1.85 &    0.26 \\
OGLE-TR-58 &    1.75 &    0.26 \\
OGLE-TR-59 &    1.15 &    0.16 \\
\hline
}

Table~2 indicates that in several cases the new transiting companions are 
of Jupiter size (${<1.6~R_{\rm Jup}}$). They may be planets, or brown dwarfs 
or low-luminosity M-type stars. The spectroscopic follow-up will provide 
necessary clarification. 

The most intriguing object within this group is OGLE-TR-56. Table~2 indicates 
that the size of the transiting companion is smaller than the Saturn size if 
its transit is central. However, even for inclination as low as ${i=83\arcd}$, 
its radius is still equal to one Jupiter radius. This object is certainly very 
small. The color of the primary star indicates a star of smaller size 
than the Sun consistent with Table~2. If confirmed spectroscopically to be a 
planet, OGLE-TR-56 would have the shortest orbital period (1.21190~days) and 
would be one of the smallest extrasolar planets. 

Alternatively the transiting object in OGLE-TR-56 might be a brown dwarf. 
Recently, Santos \etal (2002) reported possible detection of a brown dwarf 
around HD~41004 with a period of only 1.3 days. Also, possible detection 
of a brown dwarf in the short period Hyades binary system RHy~403 was 
reported by Reid and Mahoney (2000).  OGLE-TR-56 could be another 
example indicating that the so called ``brown dwarf desert'' -- lack of brown 
dwarfs with short periods on small orbits, is not that empty as generally 
thought. 

The least interesting alternative would be a blend of a regular Galactic bulge 
eclipsing star with a bright disk star which could mimic the low depth 
transit. However, this case is not very likely -- the shape of the transit 
light curve, namely short time of descending and ascending branches of the 
transit compared to long phase of totality, makes this explanation less 
probable. 

It is worth adding here that two other most promising planetary transit
objects: OGLE-TR-40 and OGLE-TR-10, discovered in the original paper
(Udalski \etal 2002), were observed for expected transits on a few nights
in June 2002. The main goal of these observations was the improvement of
ephemerides that due to short time baseline in 2001 could be inaccurate
after longer period of time. For both objects a few transits were
covered and the new best ephemerides, based now on more than one hundred
orbital cycles, are:
\vspace*{-7pt}
$${\rm HJD}_{\rm min} = 2452060.03667 + 3.43079\cdot E \eqno({\rm OGLE-TR-40})$$
\vspace*{-15pt}
$${\rm HJD}_{\rm min} = 2452070.21900 + 3.10140\cdot E \eqno({\rm OGLE-TR-10})$$
The photometric data of new OGLE-III transit objects  presented in this
paper are available in the electronic form from the OGLE archive: 
\vskip3pt
\centerline{\it http://www.astrouw.edu.pl/\~{}ogle} 
\vskip3pt
\centerline{\it ftp://ftp.astrouw.edu.pl/ogle/ogle3/transits/new$\_$2001$\_$transits}

\vskip3pt
\noindent
or its US mirror

\centerline{\it http://bulge.princeton.edu/\~{}ogle}
\vskip3pt
\centerline{\it ftp://bulge.princeton.edu/ogle/ogle3/transits/new$\_$2001$\_$transits}

\Acknow{We would like to thank Prof.\ B.\ Paczy{\'n}ski for many
interesting discussions and comments. 
The paper was partly supported by the  Polish KBN grant 2P03D01418 to 
M.\ Kubiak. Partial support to the OGLE project was provided with the NSF  
grant AST-9820314 and NASA grant NAG5-12212 to B.~Paczy\'nski.}

\end{document}